\documentclass[12pt]{iopart}
\usepackage{epsfig,graphicx,amssymb,color,bm,cite}

\newcommand{\ffref}[1]{Fig.\,\ref{#1}}

\begin{document}

\title{Optomechanical Force Sensor in Non-Markovian Regime}
\author{Wen-Zhao Zhang$^{1}$, Yan Han$^{2}$, Biao Xiong$^{1}$, Ling Zhou$%
^{1,*}$}

\begin{abstract}
The optomechanical force sensor in non-Markovian environment for a
mechanical oscillator is presented. By performing homodyne detection, we
obtain a generally expression for the output signal. It is shown that the
weak force detection is sensitive to the non-Markovian environment. The
additional noise can be reduced and the mechanical sensitivity can be obviously amplified in resolved sideband regime comparing to the Markovian condition without using assistant system or squeezing. Our results provide a
promising platform for improving the sensitivity of weak force
ultrasensitive detection.
\end{abstract}

\address{$^{1}$ School of Physics and Optoelectronic Technology, Dalian
University of Technology, Dalian 116024, People's Republic of China} %
\address{$^{2}$ School of Physics and Optoelectronic Technology, Taiyuan
University of Technology, Taiyuan 030024, People's Republic of China} %
\address{$^{*}$ Author to whom any correspondence should be addressed.} %
\ead{zhlhxn@dlut.edu.cn} \vspace{10pt}
\begin{indented}
\item[]February 2016
\end{indented}

%
%
%
%
%

\section{Introduction}

Optomechaical systems provide us a platform for high precision measurements
including ultra-sensitive force detection \cite{PhysRevA.89.023848}, small
quantities of adsorbed mass detection \cite{Kolkowitz2012} and
low-reflectivity object detection \cite{PhysRevLett.114.080503}. Such
systems exploit the huge susceptibility around the resonance frequency of
oscillators with excellent mechanical quality factor $Q_{m}$, combined with
high-sensitivity interferometric measurements \cite{PhysRevA.89.023848,QO}.
The photon shot noise in the optomechanical systems will broaden the optical
response spectrum and finally affect the sensitivity of detection during the
frequency measurement \cite{PhysRevLett.115.243603,PhysRevLett.113.151102},
which means that the shot noise should be reduced. However, reduced shot
noise would increase quantum back-action noise force due to the opposite
scalings with the optical field intensity \cite{RevModPhys.86.1391}. Many
schemes have been proposed to optimally compromise between photon shot noise
and quantum back-action \cite{srep31095}, which leads to the standard
quantum limit (SQL) in weak force sensing \cite{RevModPhys.82.1155,srep11701}%
. Various approach beyond-SQL measurements have been proposed \cite%
{PhysRevA.82.033811,Teufel2009,PhysRevA.92.043817,PhysRevA.90.043848,PhysRevA.73.033819}%
, including optical squeezing in the optomechanical system \cite%
{PhysRevA.90.043848,PhysRevLett.115.243603}, atomic assistance in a separate
cavity \cite{PhysRevA.92.043817}, mechanical modification by the light \cite%
{PhysRevA.73.033819}, and so on.
Up to now, most of the measurement schemes
are based on the Born-Markov approximation. The noise effect from a
structured bath for optomechanical measurement is still not discussed. On
the other hand, how to improve detecting precision with a structured bath is
also unresolved. Thus, investigation measurement noise under a structured
environment is a practical requirement for the further development of high
precision measurements.

Generally speaking, the quantal consideration of thermal noise of the
optomechanical measurement system can be adequate described as a movable
mirror undergoing quantum Brownian motion with the coupling through the
reservoir momentum \cite{PhysRevA.63.023812}. The dynamics of this system
are a non-Markovian process essentially. Since the non-Markovian environment
exhibits memory effect \cite%
{PhysRevA.81.052105,srep23678,PhysRevA.93.063853,PhysRevE.90.022122,PhysRevA.94.012334}
which can be used to store quantum information \cite{PhysRevA.81.052105}, to
generate and protect entanglement \cite{NJP.17.033038,srep23678} and to
enhance the side-band cooling effect \cite%
{PhysRevA.93.063853,PhysRevLett.116.183602}, it might be benefit for high
precision measurements due to the same requirements of quantum behavior
protection. Most recently, a kind of non-Markovian environment for
mechanical oscillator had been designed, in which spectrum density of the
environment was detected \cite{Groblacher2015}, which make it possible to
detect the weak force under a structured environment. With this
consideration in mind, we investigate the detection property based on an
elementary optomechanical system where the mechanical oscillator is coupled
to a non-Markovian reservoir, while the bath of the cavity is a Markovian
environment so as to output the signal of oscillator through the cavity.

In this paper, we introduce a non-Markovian environment for the mechanical
oscillator and obtain the solution of output signal under homodyne
detection. Then we study the sensibility and additional noise of an
optomechanical weak force detection system with different spectrum densities
$\mathcal{J}(\omega )$ including that of Markovian condition. We find that
some environments with super-Ohmic spectrum or experimental cut-off spectrum %
\cite{Groblacher2015} do have obvious enhanced sensibility comparing with
that under Markovian condition. Furthermore, we can greatly reduce the
additional noise even in the unsolved sideband regime.

\section{Model}

\begin{figure}[t]
\centering 
\includegraphics[width=8cm]{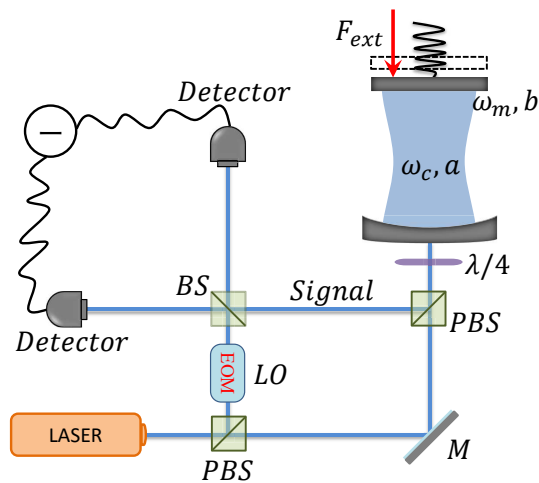}
\caption{(color online) Schematic of the system with homodyne detection. The
local oscillator (LO) is phase-modulated with an electro-optical modulator
(EOM). The monitored system is composed of a optomechanical cavity and a
general non-Markovian reservoir of the mechanical oscillator.}
\label{fig1}
\end{figure}

We consider a typical optomechanical system where the frequency of the
cavity and the mechanical resonator are $\omega _{c}$ and $\omega _{m}$,
respectively.
The weak force is sensed by the mechanical oscillator, and the environment noise simultaneously exerts a stochastic force to the oscillator.
In order to detect the signal force, we assume that the mechanical oscillator is coupled to a non-Markovian reservoir so as to decrease stochastic force. Considering the feasibility, a Markovian environment for optical mode is easy to output the optical signal to perform homodyne detection in experiment.
Therefore, we consider the optical mode in Markovian regime.
As shown in Fig.~\ref{fig1}, the output signal can be processed in the standard homodyne detection.
The Hamiltonian of the system can be described as $H=H_{S}+H_{E}$ with
\begin{eqnarray}
H_{S} &=&\hbar \omega _{c}a^{\dag }a+\frac{1}{2}\hbar \omega
_{m}(p_{m}^{2}+q_{m}^{2})-\hbar g_{0}a^{\dag }aq_{m}+i\hbar E(a^{\dag
}e^{-i\omega _{d}t}-ae^{i\omega _{d}t}),  \nonumber  \label{Hami} \\
H_{E} &=&\sum_{k}\hbar \omega _{k}[\frac{1}{2}(p_{k}^{2}+q_{k}^{2})+\gamma
_{k}q_{k}q_{m}],
\end{eqnarray}%
where $H_{S}$ describes the cavity mode driven by a laser coupled to the
mechanical resonator via radiation pressure with the coefficient $%
g_{0}=(\omega _{c}/L)\sqrt{\hbar /2m\omega _{m}}$. And $\omega _{d}$ is the
angular frequency of the laser, and $E$ is the cavity driving strength given
by $E\equiv 2\sqrt{P\kappa _{ex}/\hbar \omega _{d}}$ with $P$ the input
power of the laser and $\kappa _{ex}$ the input rate of the cavity. The
first term of $H_{E}$ is the energy of the mechanical reservoir for the $k$%
th environmental oscillator with frequency $\omega _{k}$. The second term of
$H_{E}$ describes the coupling between the mechanical oscillator and the
reservoir with the coupling strength $\omega _{k}\gamma _{k}$ for the $k$th
environmental mode. For convenience, we take $\hbar =1$ throughout the
paper. In the rotating frame at the driving laser frequency $\omega _{d}$,
the time evolution of the system and reservoir operators in the Heisenberg
picture are \numparts
\begin{eqnarray}
\dot{a} &=&-(i\Delta _{c}+\frac{\kappa }{2})a+ig_{0}aq_{m}+E+\sqrt{\kappa }%
a_{in}, \\
\dot{q_{m}} &=&\omega _{m}p_{m}, \\
\dot{p_{m}} &=&-\omega _{m}q_{m}+g_{0}a^{\dag }a-\sum_{k}\omega _{k}\gamma
_{k}q_{k}, \\
\dot{q_{k}} &=&\omega _{k}p_{k}, \\
\dot{p_{k}} &=&-\omega _{k}q_{k}-\omega _{k}\gamma _{k}q_{m},
\end{eqnarray}%
\label{tleq} \endnumparts where $\Delta _{c}=\omega _{c}-\omega _{d}$,
$\kappa $ and $a_{in}$ denote the dissipation rate and noise operator of the cavity, respectively.
The autocorrelation function of the vacuum noise is $\langle a_{in}(t)a_{in}^{\dag }(\tau )\rangle =\delta
(t-\tau )$.
Solving Eqs.~(\ref{tleq}d) and (\ref{tleq}e), we have
\begin{eqnarray}
q_{k}(t) &=&q_{k}(0)\cos (\omega _{k}t)+p_{k}(0)\sin (\omega _{k}t)-\omega
_{k}\gamma _{k}\int_{0}^{t}d\tau q_{m}(\tau )\sin [\omega _{k}(t-\tau )].
\nonumber \\
&&
\end{eqnarray}%
Substituting it into Eq.~(\ref{tleq}c),
\[
\dot{p_{m}}=-\omega _{m}q_{m}+g_{0}a^{\dag }a+\int_{0}^{t}d\tau f(t-\tau
)q_{m}(\tau )+F_{in},
\]%
where $f(t)=\sum_{k}\omega _{k}^{2}\gamma _{k}^{2}\sin \omega _{k}t=\int
\frac{d\omega }{2\pi }\mathcal{J}(\omega )\sin \omega t$. $%
F_{in}=F_{ext}+\xi (t)$, here $\xi (t)=-\sum_{k}\omega _{k}\gamma
_{k}[q_{k}(0)\cos (\omega _{k}t)+p_{k}(0)\sin (\omega _{k}t)]$ is the
input-noise of the oscillator, which depend on the initial states of the
reservoir. In Markovian regime this term is usually written as $\sqrt{\gamma
_{m}}F_{th}$, where $\gamma _{m}$ is the dissipation rate of the mechanics,
and $F_{th}$ is the noise operator. $F_{ext}$ is the external forces to be
measured \cite{PhysRevA.90.043848,PhysRevA.92.043817}, which can be a
accelerated mass \cite{apl.104.22}, magnetostrictive material \cite%
{PhysRevLett.108.120801}, atomic force \cite{OE.20.018268} or gravitational
waves \cite{PhysRevLett.115.211104}. Currently, most experimental
realizations of cavity optomechanics are still in the single-photon weak
coupling with strong driving condition \cite%
{Groblacher2009,PhysRevLett.110.233602,PhysRevE.93.062221,Teufel2011}. Under
this condition, we can linearize the equations of motion around the steady
state with $p_{m}\rightarrow p_{m}+p_{0}$, $q_{m}\rightarrow q_{m}+q_{0}$, $%
a\rightarrow \alpha +a$, here $p_{0}\equiv \langle p_{m}\rangle $, $%
q_{0}\equiv \langle q_{m}\rangle $ and $\alpha \equiv \langle a\rangle $.
Neglecting the nonlinear term, we write the linearized quantum Langevin
equations as
\begin{eqnarray}
\dot{a} &=&-(i\Delta _{c}^{\prime }+\frac{\kappa }{2})a+iGq_{m}+\sqrt{\kappa
}a_{in}, \\
\dot{q_{m}} &=&\omega _{m}p_{m},  \nonumber \\
\dot{p_{m}} &=&-\omega _{m}q_{m}+\int_{0}^{t}d\tau f(t-\tau )q_{m}(\tau
)+G^{\ast }a+Ga^{\dag }+F_{in},  \nonumber
\end{eqnarray}%
where $G=\alpha g_{0}$ is the linearized coupling strength, $\Delta
_{c}^{\prime }=\Delta _{c}-g_{0}q_{0}$ denotes the effective detuning of the
cavity. In order to solve the dynamics of the system and to find the input
noise sources, we now switch into the frequency domain by introducing the
Fourier transform operator $O(\omega )=\frac{1}{\sqrt{2\pi }}\int
dtO(t)e^{i\omega t}$ and obtain
\begin{eqnarray}
a(\omega ) &=&\chi _{c}(\omega )[iGq_{m}(\omega )+\sqrt{\kappa }%
a_{in}(\omega )],  \label{eqw} \\
q_{m}(\omega ) &=&\chi _{m}(\omega )[G^{\ast }a(\omega )+Ga^{\dag }(-\omega
)+F_{in}(\omega )],  \nonumber
\end{eqnarray}%
where $\chi _{c}\equiv \lbrack \kappa /2-i(\omega -\Delta _{c}^{\prime
})]^{-1}$ and $\chi _{m}\equiv -\omega _{m}/[(\omega ^{2}-\omega
_{m}^{2})+\omega _{m}\Sigma (\omega )]$ are susceptibilities of cavity and
mechanical oscillator with $\Sigma (\omega )=\mathcal{P}\int d\omega
^{\prime }\frac{\omega ^{\prime }\mathcal{J}(\omega ^{\prime })}{(\omega
^{2}-\omega ^{\prime }{}^{2})}\mp i\pi \frac{\theta(\omega)\mathcal{J}(\omega )-\theta(-\omega)\mathcal{J}%
(-\omega )}{2}$ the Laplace transform of the self-energy correction \cite%
{PhysRevLett.109.170402,srep23678}, where $\theta(\omega)$ is a step function.
Here $\chi _{m}$ denotes the effect of
the mechanical bath which depends on the spectrum density $\mathcal{J}%
(\omega )$. The commonly used ohmic-type spectral density of the form $%
\mathcal{J}(\omega )=\eta \omega (\frac{\omega }{\omega _{0}})^{s-1}e^{-%
\frac{\omega }{\omega _{0}}}$ \cite%
{RevModPhys.59.1,PhysRevLett.109.170402,RevModPhys.86.1203}, where $\eta $
is the strength of system-bath coupling, and $\omega _{0}$ is the cut-off
frequency. The exponent $s$ is a real number that determines the $\omega $
dependence of $\mathcal{J}(\omega )$ in the low-frequency region. The baths
with $0<s<1$, $s=1$, and $s>1$ are termed as ``sub-Ohmic'', ``Ohmic''\ and
``super-Ohmic''\ baths, respectively. In Markovian condition $\chi
_{m}=-\omega _{m}/[(\omega ^{2}-\omega _{m}^{2})+i\gamma _{m}\omega ]$,
where $\gamma _{m}$ is the dumping rate of the mechanical oscillator.
Similarly, in non-Markovian regime we can also define the equivalent
dissipation rate $\gamma _{eff}$ which depends on the spectrum density $%
\mathcal{J}(\omega )$. Solving Eq.~(\ref{eqw}), we have
\begin{equation}
q_{m}(\omega )=\frac{G^{\ast }\chi _{c}\sqrt{\kappa }a_{in}(\omega )+G\chi
_{c}^{\prime }\sqrt{\kappa }a_{in}^{\dag }(-\omega )+F_{in}(\omega )}{\chi
_{m}^{-1}-i|G|^{2}(\chi _{c}-\chi _{c}^{\prime })},  \label{eqq}
\end{equation}%
where $\chi _{c}^{\prime }=[\kappa /2-i(\omega +\Delta _{c}^{\prime })]^{-1}$%
. In Eq.~(\ref{eqq}) the coordinate of mechanical operator in frequency
domain is composed of two parts. The one term is proportion to the input
field of the cavity through the radiation pressure coupling with the
coefficient $G$. The other term $F_{in}(\omega )$ is resulted from the bath
of the oscillator and external force. If we neglect the effect from the
cavity, we can rewrite Eq.~(\ref{eqq}) as $q_{m}(\omega )=\chi _{m}(\omega
)F_{in}$. There is an obvious positive correlation between the external
force and the position spectrum of the oscillator. For weak force detection,
we need a large susceptibility $\chi _{m}(\omega )$ to magnify the weak
signal $F_{ext}$. On the other hand, the thermal noise from the environment
should be reduced, because the noise can be coequally amplified with the
detecting signal by the system. In Markovian regime, the two requirements
will demand a high mechanical quality factor and low bath temperature \cite%
{PhysRevA.89.023848}. But it is more complex in non-Markovian condition, $%
\chi _{m}(\omega )$ totally depends on the self-energy correction $\sum
(\omega )$ which is a frequency dependent parameter up to the structure of
the bath. We will make a specific discuss in the follow section.\newline
\indent It is hard for us to direct detect the oscillator experimentally.
But the signal from the external force can be output and enhanced by the
cavity through the optomechanical interaction. Usually, we use the output
photon from the optomechanical cavity as an indirect information carrier.
Under Markovian regime for the optical field, we can use the standard input-output relation $O_{out}=\sqrt{\kappa }O-O_{in}$.
Considering a hommodyne measurement shown in Fig.~\ref{fig1}, we have the
signal
\begin{eqnarray}
M_{out}(\omega ) &=&i[a_{out}^{\dag }(-\omega )e^{-i\theta }-a_{out}(\omega
)e^{i\theta }]  \label{eqout} \\
&=&A(\omega )a_{in}(\omega )+B(\omega )a_{in}^{\dag }(-\omega )+C(\omega
)F_{in},  \nonumber
\end{eqnarray}%
where
\begin{eqnarray}
A(\omega ) &=&\frac{4e^{i\theta }\kappa G^{\ast 2}\chi _{m}-ie^{-i\theta
}[D(4|G|^{2}\chi _{m}-D)+4\omega ^{2}]}{\sqrt{2}[4\Delta _{c}^{\prime
}(\Delta _{c}^{\prime }-2|G|^{2}\chi _{m})+(\kappa -2i\omega )^{2}]},
\nonumber \\
B(\omega ) &=&\frac{4e^{-i\theta }\kappa G^{2}\chi _{m}+ie^{i\theta
}[D^{\ast }(4|G|^{2}\chi _{m}-D^{\ast })+4\omega ^{2}]}{\sqrt{2}[4\Delta
_{c}^{\prime }(\Delta _{c}^{\prime }-2|G|^{2}\chi _{m})+(\kappa -2i\omega
)^{2}]},  \nonumber \\
C(\omega ) &=&\frac{2i\sqrt{\kappa }\chi _{m}[e^{i\theta }G^{\ast }(D^{\ast
}-2\omega )-e^{-i\theta }G(D+2\omega )]}{\sqrt{2}[4\Delta _{c}^{\prime
}(\Delta _{c}^{\prime }-2|G|^{2}\chi _{m})+(\kappa -2i\omega )^{2}]},
\nonumber \\
&&
\end{eqnarray}%
with $D=2\Delta _{c}^{\prime }+i\kappa $, and the phase $\theta $ is
introduced and can be optimized to enhance the sensitivity of the weak force
detection \cite{PhysRevA.90.043848}. To obtain the relationship between the
detecting force and the output signal, we can rewrite Eq.~(\ref{eqout}) as
\begin{equation}
\frac{M_{out}(\omega )}{C(\omega )}=\frac{A(\omega )}{C(\omega )}%
a_{in}(\omega )+\frac{B(\omega )}{C(\omega )}a_{in}^{\dag }(-\omega )+F_{in}.
\end{equation}%
Considering $F_{in}=F_{ext}+\xi (\omega )$ and defining $M_{out}(\omega
)/C(\omega )=F_{ext}+F_{add}(\omega )$, then we have
\begin{equation}
F_{add}=\xi (\omega )+\frac{A(\omega )}{C(\omega )}a_{in}(\omega )+\frac{%
B(\omega )}{C(\omega )}a_{in}^{\dag }(-\omega ),
\end{equation}%
where $F_{add}$ is the additional noise of the detecting force. The first
term denotes the thermal noise operator of the mechanical environment, and
the second and third term denote the input noise of the cavity. From the
general definition of the noise spectrum, we have
\begin{equation}
S_{add}(\omega )=\frac{1}{2}[S_{FF}(\omega )+S_{FF}(-\omega )],
\end{equation}%
where $S_{FF}(\omega )=\int d\omega ^{\prime }\langle F_{add}(\omega
)F_{add}(\omega ^{\prime })\rangle $. Here we assume that any two parts
initially have no correlation. The vacuum radiation input noise $a_{in}$
satisfy $\delta $-correlation function. The additional noise spectrum
density becomes
\begin{equation}
S_{add}(\omega )=S_{\xi \xi }(\omega )+\frac{|A(\omega )|^{2}+|B(\omega
)|^{2}}{2|C(\omega )|^{2}},  \label{eqadd}
\end{equation}%
where $S_{\xi \xi }(\omega )$ is thermal noise with the structured bath
which does not depended only on the bath temperature but also on the
spectrum density $\mathcal{J}(\omega )$. For simplicity, we choose $\theta =0
$, then Eq.~(\ref{eqadd}) can be rewritten as
\begin{eqnarray}  \label{eqsadd}
S_{add}(\omega) &=&S_{\xi\xi}(\omega) +\frac{|P(\omega)|^2+|Q(\omega)|^2}{8\kappa|G^*D^*-GD-2\omega(G+G^*)|^2}.
\end{eqnarray}
where
\begin{eqnarray}
P(\omega ) &=&4(\kappa G^{\ast 2}-iD|G|^{2})+\frac{i(D^{2}-4\omega ^{2})}{%
\chi _{m}}  \nonumber  \label{eqpq} \\
Q(\omega ) &=&4(\kappa G^{2}+iD^{\ast }|G|^{2})-\frac{i(D^{\ast 2}-4\omega
^{2})}{\chi _{m}}
\end{eqnarray}%
$S_{add}(\omega )$ is also defined as a effective force noise spectral
density to evaluate the sensitivity to the external force \cite%
{PhysRevA.90.043848}. We will show that the effective force noise can be
reduced by engineering the environment.

\section{The mechanical susceptibility and thermal correlation with a
structured environment}

\label{sec3}

\begin{figure}[tbp]
\centering\includegraphics[width=12cm]{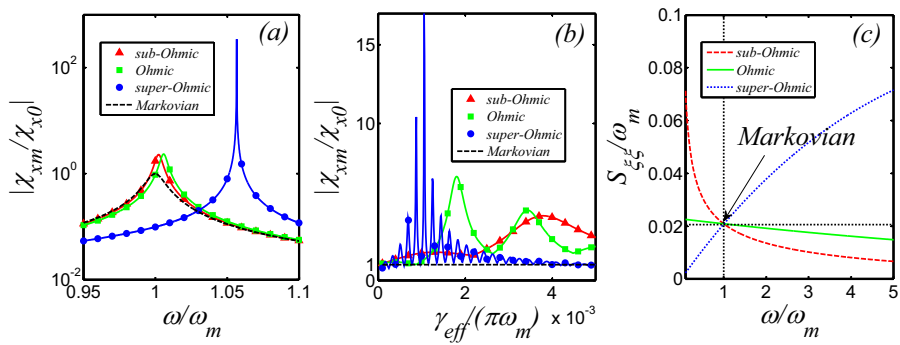}
\caption{(Color online) (a) The ratio of mechanical sensibility $\protect%
\chi _{xm}/\protect\chi _{x0}$ as a function of $\protect\omega$ with
Markovian condition and ohmic-type spectrum. $s=0.5,1,2$ for three kinds of
ohmic-type spectrum, respectively, the equivalent dumpling rate $\protect%
\gamma _{eff}/\protect\omega _{m}=\protect\pi \times 10^{-3}$. (b) The maximal ratio
of mechanical sensibility $\protect\chi _{xm}/\protect\chi _{x0}$ as a
function of equivalent dumpling rate $\protect\gamma _{eff}$ for different spectrum. (c)
Thermal noise with three kind of ohmic-type spectrum density. In Markovian
regime, the noise exist only at the frequency $\protect\omega _{m}$, the
directrix is plotted by black dashed line, while $s=0.5$ for sub-Ohmic
spectrum, $s=1$ for Ohmic spectrum, $s=2$ for super-Ohmic spectrum. The
equivalent dissipation $\protect\gamma _{eff}=\protect\pi \times 10^{-3}%
\protect\omega _{m}$. Other parameters are, the oscillator frequency $%
\protect\omega _{m}=10^{6}Hz$, the bath temperature $T=1mK$, the cut-off
frequency $\protect\omega _{0}/\protect\omega _{m}=10$.}
\label{fig2}
\end{figure}

Before we investigate the additional noise spectrum, we first analyze the
effect of the ability of amplification $\chi _{m}(\omega )$ and thermal
noise spectrum $S_{\xi \xi }(\omega )$ under non-Markovian environment. As
we have mentioned in last section, the sensibility of the mechanical
oscillator for the weak force ultrasensitive detection in optomechanical
system is determined by the quantity $\chi _{m}(\omega )$ and has been
widely discussed in Markovian regime\cite%
{PhysRevLett.99.110801,PhysRevLett.96.173901}. In non-Markovian regime, $%
\chi _{m}(\omega )$ is a spectrum depended parameter.
According to \eref{eqq}, considering the effect of cavity, we now present the character of mechanical susceptibility by plotting $\chi _{xm}(\omega)/\chi_{x0}$ as a function of $\omega $ with the commonly used ohmic-type spectrum
in Fig.~\ref{fig2}a and b, where $\chi_{xm}^{-1}=\chi_{m}^{-1}-i|G|^{2}(\chi _{c}-\chi _{c}^{\prime })$ denote the mechanical sensitivity, $\chi _{x0}^{-1}=-i\gamma _{eff}-i|G|^{2}[\chi _{c}(\omega_m)-\chi _{c}^{\prime }(\omega_m)]$ is optimal mechanical sensitivity in Markovian regime.
As shown in Fig.~\ref{fig2}a,
it is clearly seen that the maximal sensitive frequency area is around the
oscillator frequency $\omega _{m}$ in Markovian regime, which is consistent
with the generally results in weak force detection in Ref.~\cite%
{PhysRevA.89.023848}. For ohmic-type spectrum, super-Ohmic environment could
provide an obviously amplification for susceptibility of detection.
We also notice that, a structured bath will cause a frequency
displacement of the maximal susceptibility, because $\omega _{m}$ is
substituted by the effective frequency $\omega _{eff}\approx \sqrt{%
\omega_{m}[\omega _{m}+\mathcal{P}\int d\omega ^{\prime }\frac{\omega
^{\prime }\mathcal{J}(\omega ^{\prime })}{(\omega _{m}^{2}-\omega ^{\prime
2})}]}$ for a structured reservoir \cite{srep23678}. And the optimal
detection area should be on resonance with this effective frequency
(Resonance Amplification).

In Fig.~\ref{fig2}b, we plot the maximal ratio of mechanical sensitivity $\chi _{xm}(\omega)/\chi_{x0}$ with different environment as a function of dumpling rate $\gamma _{eff}$.
Here $\gamma _{eff}$ is a spectrum depended parameter which describes the dissipation strength of the structured bath, through the inverse Laplace transform of $\Sigma (\omega )$ we have $\gamma_{eff}\approx\pi \mathcal{J}(\omega _{m})$ \cite{PhysRevA.91.062121}.
Under this condition, $\gamma _{eff}$ proportion to the system-bath coupling strength $\eta$.
It is shown that, for ohmic-type spectrum, the maximal ratio of mechanical sensitivity $\chi _{xm}(\omega)/\chi_{x0}$ exhibits vibration behavior with the increase of effective dissipation rate $\gamma_{eff}$.
The maximal sensitivity ratio of the system will reach the peak value at some specific effective dissipation rate $\gamma_{eff}$, and does not require the system-environment coupling factor $\eta$ is too strong.
Thus, we can significantly enhance the sensitivity with a structured environment, and the corresponding detection frequency should also be modulated.

The thermal noise $S_{\xi \xi }(\omega )$ as background noise negatively
affects the weak force detection. In order to improve the precision of the
weak force detection, we should reduce the effect result from the thermal
noise of the bath of the oscillator. We consider a movable mirror undergoing
quantum Brownian motion reservoir. The thermal-noise spectral density is
defined as $S_{\xi \xi }(\omega )\equiv \int_{-\infty }^{+\infty
}dte^{i\omega t}\langle \xi (t)\xi (0)\rangle $. Considering the structure
of the environment, we have
\begin{eqnarray}
S_{\xi \xi }(\omega ) &=&\int_{-\infty }^{+\infty }dte^{i\omega t}
\int_{0}^{\infty }d\omega ^{\prime }\mathcal{J}(\omega ^{\prime
})[n_{th}(\omega ^{\prime })\cos (\omega ^{\prime }t)+\frac{1}{2}%
e^{-i\omega^{\prime }t}],  \label{tmn}
\end{eqnarray}%
where $n_{th}(\omega )=(e^{\frac{\hbar \omega }{k_{B}T}}-1)^{-1}$ is
phononic distribution function of the reservoir. $\mathcal{J}(\omega )$
describes the character of the reservoir. For Born-Markov approximation
where the system-reservoir coupling rate is weak and the interaction time is
short enough, the environment can be described as a flat spectrum and the
integral for $\omega $ is a delta function of time; therefore, the
environment present no memory effect for the system, i.e. $S_{\xi \xi
}=\gamma _{m}n_{th}(\omega _{m})$, where $\gamma _{m}$ is the dumping rate
of the mechanical oscillator, $n_{th}(\omega _{m})$ describes the equivalent
thermal occupation which is independent of the environment frequency.

As shown in Fig.~\ref{fig2}c, we plot the thermal noise spectral density for
Markovian and non-Markovian environment as a function of $\omega$. For the
exponent $s=0.5,1,2$ the corresponding coupling strength of system-bath for
sub-Ohmic, Ohmic and super-Ohmic are $\eta _{0.5}=5.5\times 10^{-3}$, $\eta
_{1}=1.2\times 10^{-2}$ and $\eta_{2}=6.1\times 10^{-2}$, respectively, when
we choose the same equivalent dumping rate $\gamma _{eff}=\pi \times
10^{-3}\omega _{m}$ as that for Markovian condition. For a fair comparison,
the other parameters are also selected the same for different structured
bath. From Fig.~\ref{fig2}c, we see that different structure of bath will
cause different distribution of thermal excitation. But, around frequency $%
\omega_{m}$, $S_{\xi\xi }\approx 0.02$ for different reservoir.
While the effective frequency $\omega _{eff}$ just shifts slightly, we can
reasonably ignore the thermal noise $S_{\xi \xi }$ because one can observe
that $S_{\xi\xi }$ is below 0.025 around $\omega _{m}$. Under Markovian
reservoir, according to Eq.~(\ref{tmn}), for the common used detection
frequency area $\omega _{m}$, the noise $S_{\xi \xi }\approx \gamma _{eff}%
\frac{k_{b}T}{\hbar \omega _{m}}=k_{b}T/(\hbar Q_{eff})$. Thus we can reduce
the thermal noise by cooling down the system \cite{PhysRevA.93.063853} or
improve the effective mechanical quality factor $Q_{eff}$ directly.
According to the experiment parameters in nano-mechanical system \cite{Teufel2009}, where $\omega _{m}=2\pi \times 1.04 \textrm{MHz}$, mechanical
quality factor $Q_{m}=6.2\times 10^{5}$ and the environment temperature $%
T=77\textrm{mK}$, the thermal noise $S_{\xi \xi }/\omega _{m}\ll 1$, which
means that we can ignore the thermal noise for weak force detection under
current experimental conditions \cite{PhysRevLett.113.151102}.

\section{The additional noise with a structured environment}

For weak force detection, in addition to a high sensitivity, good linearity
and high response speed, we expect to reduce additional noise, which is also
widely used as a detection waveband, such as weak force detection through
OMIT \cite{PhysRevA.93.023802}, microwave quantum illumination by
optomechanical system \cite{PhysRevLett.114.080503}, gravitational-wave
detectors with unstable optomechanical filters \cite{PhysRevLett.115.211104}%
. Now, we show that under certain environment we can obtain high sensitivity
and reduced additional noise.\newline
\indent Recently, a spectral density of mechanical environment had been
detected experimentally through the emitted light of miro-optomechanical
system \cite{Groblacher2015}. The demonstration device consists of a thick
layer of $Si_{3}N_{4}$ with a high-reflectivity mirror pad in its centre as
a mechanically moving end mirror in a Fabry-P\'{e}rot cavity where the
spectral density is described by $\mathcal{J}(\omega )=C\omega ^{k}$ with $%
C>0$ and $k=-2.30\pm 1.05$. The region of $\omega $ satisfies $\omega \in
\lbrack \omega _{min},\omega _{max}]$ centred around mechanical resonance
frequency $\omega _{m}=914kHz$. Here $\omega _{min}=885kHz$ and $\omega
_{max}=945kHz$, the corresponding bandwidth $\Gamma \approx 0.07\omega _{m}$%
. Employing this cut-off experimental spectral density $\mathcal{J}(\omega
)=C\omega ^{k}$, where $C=\mathcal{J}(\omega _{m})/\omega _{m}^{k}$, and we
choose the bandwidth $\Gamma _{m}=0.2\omega _{m}$, exponent $k=-2$.
\begin{figure}[tbph]
\vspace{0.5cm} \centering 
\includegraphics[width=13cm]{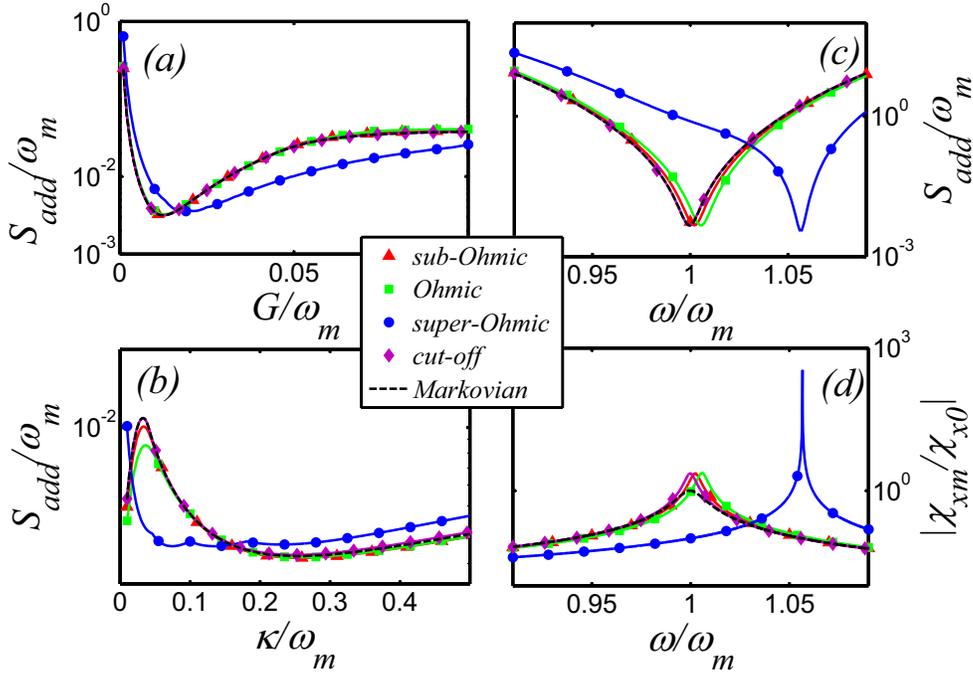}
\caption{(Color online) (a), (b) and (c) describe the optimal additional
noise $S_{add}$ as a function of linearized coupling rate $G$, dispassion
rate $\protect\kappa$ and frequency $\protect\omega$, respectively. (a) The
dispassion rate $\protect\kappa/\protect\omega_m=0.1$. (b) The linearized
coupling rate $G/\protect\omega_m=0.02$. (c) The dispassion rate $\protect%
\kappa/\protect\omega_m=0.1$ and linearized coupling rate $G/\protect\omega%
_m=0.02$. (d) The ratio of mechanical susceptibility $\protect\chi_{xm}/\protect%
\chi_{x0}$ as a function of $\protect\omega$ with different spectrum. The
equivalent dumpling rate $\protect\gamma_{eff}/\protect\omega_m=\protect\pi %
\times 10^{-3}$. Other parameters are same with Fig.~\ref{fig2}.}
\label{fig3}
\end{figure}

We plot the additional noise and the susceptibility for the different types
of environment of the mechanical oscillator in Fig.\,\ref{fig3}, where we
reasonably ignore thermal noise $S_{\xi \xi }$ around $\omega _{m}$\cite%
{PhysRevLett.113.151102} in Eq.\,(\ref{eqsadd}) according to the conclusion
in Sec.~\ref{sec3}. As shown in Fig.~\ref{fig3}a, we plot the optimal
additional noise $S_{add}$ as a function of linearized coupling rate $G$.
It is obvious that for different spectrum, there are minimum values of $S_{add}$ at certain value of $G$.
In addition to the super-Ohmic spectrum, the evolution trend of the curve with the coupling rate $G$ is almost the same.
When the coupling rate $G/\omega_m$ is less than $0.017$, the additional noise of the super-Ohmic spectrum is larger than that of other ones.
On the contrary, the additional noise of the super-Ohmic spectrum will be less than that of other spectrums.
In the large $G$ scale, the additional noise is independent of the structure of the
environment. Under this region, the additional
noise mainly governed by the vacuum fluctuations of the cavity through
optomechanical interaction, and the noise from the mechanical environment
can be ignored. Thus, in order to reduce the additional noise the driving
strength of the cavity should not be too strong.

In Fig.~\ref{fig3}b, we plot the optimal additional noise $S_{add}$ as a
function of damping rate $\kappa $ which can be adjusted by Q-technology in
the measurement \cite{PhysRevLett.110.153606}.
As shown in Fig.~\ref{fig3}b, in addition to the super-Ohmic spectrum, evolution curve of the additional noise with the dissipation rate $\kappa$ tends to be consistent.
There is a peak value of the additional noise at the specific dissipation rate $\kappa$.
With the increase of the dissipation rate, the additional noise decreases first and then increases gradually.
For the super-Ohmic spectrum, when the dissipation rate $\kappa/\omega_m$ is less than a specific value $0.16$, the additional noise is much smaller than that of other structures.
With the increase of the dissipation rate, the additional noise of the super-Ohmic spectrum will be larger than that of other ones.
In resolved sideband regime, the additional noise of the system for different spectral structures can be maintained at a low level $10^{-3}$.
That is to say, in the weak dissipation region, the super-Ohmic spectrum can effectively reduce the additional noise.
However, there is no obvious effect on reducing the additional noise for other spectral structures under the same effective dissipation $\gamma_{eff}$.

Employing the optimized parameters based on Figs.~\ref{fig3}a and %
\ref{fig3}b, we plot the additional detection noise and sensitivity in frequency region shown in Figs.~\ref{fig3}c and \ref{fig3}d.
The additional noise $S_{add}/\omega_m$ can be reduced to near $10^{-3}$ at the
effective frequency $\omega _{eff}$.
The mechanical sensitivity of the system has been significantly improved for the ohmic-type spectrum.
Especially for the super-Ohmic spectrum, the sensitivity is about $10^3$ times of Markovian condition.

Comparing Fig.~\ref{fig3}c and Fig.~\ref{fig3}d, one can observe
that the frequency region with minimum detection noise is exactly the
frequency region with optimal sensibility. As shown in the numerator of the
second term of Eq.\,(\ref{eqsadd}) and Eq.\,(\ref{eqpq}), the maximal
sensibility $|\chi _{m}|$ results in the optimal $S_{add}$ (Because $D$ is
independent with frequency $\omega$). Thus, in our scheme, we can detect the
weak force with maximal sensibility and minimum additional noise.

Considering the feasibility, we can easy to adjust the rate $G/\omega _{m}$
by controlling the driving power of the cavity so as $G/\omega _{m}<0.05$
(see Fig.~\ref{fig3}a). Since the \emph{cut-off} spectrum \cite%
{Groblacher2015} has been realized in experiment, we can employ it to reduce
the addition noise even in the unsolved sideband regime which shows in Fig.~%
\ref{fig3}b. In order to maintain the coherence, the low-loss rate of the
mechanical oscillator is still needed.

\section{Force sensing in general environment}

\begin{figure}[t]
\centering \includegraphics[width=13cm]{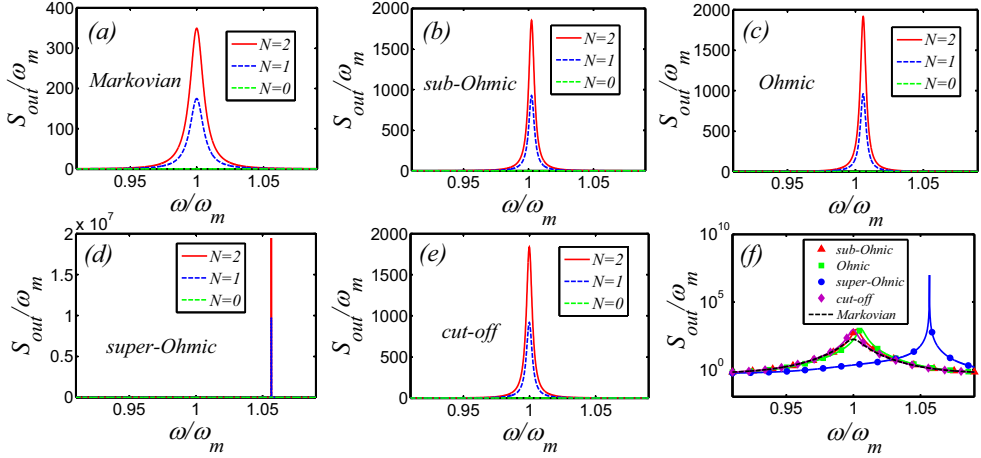}\newline
\caption{The output signal $S_{out}(\protect\omega)$ as function of
frequency $\protect\omega$ after landing the chromosomes ($N=0,1,2$) with
the bath type of (a) Markovian condition, (b) sub-Ohmic spectrum, (c) Ohmic
spectrum and (d) super-Ohmic spectrum. (e) Experimental cut-off spectrum.
(f) Comparison of different spectral structure of output signal with $N=1$.
Other parameters are same with Fig.\,%
\ref{fig2}.}
\label{fig4}
\end{figure}
\begin{figure}[tbp]
\centering \includegraphics[width=12cm]{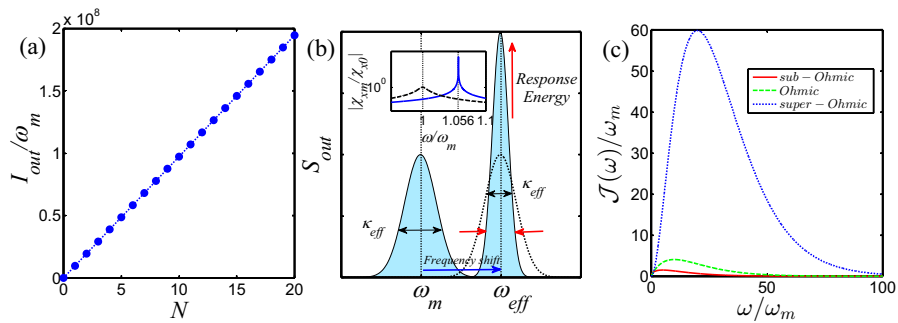}\\
\caption{(a) The linear relationship between the resonance response energy $%
I_{out}$ and the number of the chromosomes $N$ in super-Ohmic environment.
(b) Comparison of output spectrum between Markovian condition (centre around
$\protect\omega_m$) and super-Ohmic environment (centre around $\protect%
\omega_{eff}$).
(c) The spectrum of three ohmic-type environment.
The parameters are same with Fig.\,\ref{fig2}.}
\label{fig5}
\end{figure}

In order to show the advantages of the detection under non-Markovian
environment. We provide an example to measure the mass of the human
chromosome-1. The mass of one chromosome-1 molecule is about $2.7 \times
10^{-13}\textrm{g}$ \cite{apl.106.121905}. The external accretion mass will
introduce an additional frequency responded by mechanical resonator. The
mass response of the mechanical resonator can be defined as $\mathcal{R}%
=\partial\omega/\partial m$ with the typical value $\mathcal{R}=10^{21}\,%
\textrm{Hz}\cdot \textrm{g}^{-1}$ \cite{nl.8.3735,nl.6.583}. Then, we deposit a
few chromosomes onto the surface of the mechanical resonator and observe the
output signal of the system. As shown in Fig.\,\ref{fig4}, we plot the
output signal $S_{out}(\omega)$ (details see Appendix) of the optomechanical
mass sensor with differently structured environment. For fair comparison, we
choose the same effective dissipation rates as well as optomechanical cavity
parameters for different environments. From Fig.\,\ref{fig4}a to \ref{fig4}%
e, we can see that the energy of the output spectrum of the resonance
frequency increases as the number of adsorbed chromosomes increases. That is
to say, we can detect the number of the chromosomes by measuring the
strength of the resonant output energy $I_{out}=S_{out}(\omega_{eff})$.
By comparing the Markovian condition, ohmic-type spectrums and experimental cut-off spectrum in \ffref{fig5}f, we find that the detection energy response is enhanced while the noise is reduced (the bandwidth is narrowed)
for the ohmic-type and cut-off spectrum. Especially for super-Ohmic
spectrum, the energy response of mass detection is almost $5\times 10^4$ times of the
Markovian condition's.
This conclusion is similar to what we discussed in
the previous section: we can detect the weak force with maximal sensibility
and minimum additional noise in specific non-Markovian environment.

As shown in Fig.\thinspace \ref{fig5}a, there is a significant linear
relationship between the resonance response energy $I_{out}$ and the number
of the chromosomes $N$. Therefore, our scheme totally consistent with the
basic requirements of weak force detection. As shown in Fig.\thinspace \ref%
{fig5}b, by comparing two different conditions, we can see that the output
energy of the detection can be concentrated in a small area around the
effective frequency $\omega _{eff}$ due to the specific structure of the
super-Ohmic spectrum. The sensitivity or the response energy of the sensor
to the input signal in super-Ohmic environment is much higher than that in
Markovian condition, which can be seen in the subgraph of Fig.\thinspace \ref%
{fig5}b. As we have chosen the same effective dissipation rate of the
oscillator and other parameters of the optomechanical system, the noise
energy from the cavity and mechanical environment are the same. So the
signal-noise ratio of the mass sensor in super-Ohmic environment is larger
than that in Markovian condition while the bandwidth of the output spectrum
is narrowed. This process is similar to ``squeezing'' the response energy of
the input signal. We can understand the mechanism of the optimized weak
force detection in non-Markovian regime by analyzing the response of the
output field to non-Markovian environment. The mechanical oscillator is a
sensor of weak force while the external force can be regarded as a part of
the mechanical environment undoubtedly. The environment of the sensor
affects the weak force detection which can be seen in Fig.\thinspace %
\ref{fig4}. Thus, the response of the weak force detection system depended
on the coupling effect between the mechanical oscillator and it's
environment. The only different in the comparison between the Markovian and
non-Markovian condition is the characters of the environment structure. In
Born-Markov approximation, the environment be equivalent to a flat spectrum.
The effective response of the oscillator to the environment is the
combination of the average coupling effect of all bath mode and frequency
detuning between the mechanical mode and bath mode, which can be
seen in the subgraph of Fig.\thinspace \ref{fig5}b, the response coefficient
$\chi _{xm}$ is a symmetrical distribution around $\omega _{m}$. But the
system-bath response is depended on the environment spectrum $\mathcal{J}%
(\omega )$ for structured bath.

In order to understand the reason why the super-Ohmic spectrum is superior to others, we plot \ffref{fig5}c.
As shown in \ffref{fig5}c, the  $\mathcal{J}(\omega )$ distribution of super-Ohmic spectrum mainly concentrates in the low frequency region (detection region), and the contribution of the high frequency mode of the environment
can even be ignored.
But the distribution for sub-Ohmic and Ohmic spectrum are more gentle than the super-Ohmic spectrum.
Therefore, we can safely say that the super-Ohmic spectrum are more far away from the flat spectrum of Markovian environment than that for sub-Ohmic and Ohmic spectrum.
As we all know, the Markovian environment only contributes a stochastic force.
So, it is reasonable that the non-Markovian backaction can reduced the noise.
Since the super-Ohmic spectrum is the most different from the Markovian flat spectrum, it can be the  best for decreasing noise force.
Subgraph in \ffref{fig5}b clearly show that the response coefficient $\chi _{xm}$ of super-Ohmic spectrum is much higher than that for the other spectrum in the detection frequency region.
In addition, super-Ohmic spectrum of its superiority in decreasing noise \cite{Srep.5.13352} over Ohmic and sub-Ohmic spectrum had also been observed in cooling the mechanical oscillator \cite{PhysRevA.93.063853}.
Therefore, in our scheme, the output signal of the weak force detection could exhibit high response and high sharpness spectrum in non-Markovian regime, which can be implemented to improve the detection accuracy for both energy response sensor \cite{PhysRevLett.108.133601,PhysRevB.93.245407} and frequency response sensor \cite{Naik2009,apl.106.121905}.

In above discussion, the environment of the optical cavity is Markovian.
If we would like to introduce non-Markovian environment for the cavity field, the additional term
$H_{CE}=\sum_{k}[\hbar \nu _{k}b_{k}^{\dagger }b_{k}+i\hslash \mathcal{K}_{k}(b_{k}a^{\dagger }-h.c.)]$ should be added in the Hamiltonian $H$.
Insteadly, Eq.~(2a) should be substituted by another two equations, and the output relation also should be renewed. So, the problem become very complicated.
It is hard for us to directly foresee the function of the non-Markovian environment of the cavity field.
We will finish it elsewhere.

\section{Conclusion}

We investigate the weak force detection of optomechanical system in
non-Markovian regime. By solving the exact dynamics of the optomechanical
system, we obtain an general analytical result of the output signal. We have
shown that: (i) The thermal noise for weak force detection can be ignored
even under non-Markovian environment, while the susceptibility is
efficiently amplified in the effective frequency region $\omega _{eff}$.
(ii) The additional noise can be significantly reduced in super-Ohmic spectrum. The additional noise
can be maintained at a quite low level and the quantum effect can be better
protected with a structured bath by comparing with the Markovian condition in resolved sideband regime.
(iii) Employing super-Ohmic environment to reduce the additional
noise and amplification detection signal do not require the high quality of the cavity. Meanwhile,
optimized $G/\omega _{m} $ is demanded. Furthermore, we provide an example
by comparing the Markovian and non-Markovian conditions to measure the mass
of the human chromosome-1, and then we analyze the mechanism of the
optimization detection in non-Markovian regime. Instead of introducing
squeezing and improving the experiment conditions such low bath temperature
and high mechanical quality factor, our results provides another effective
way for reducing the additional noise by utilizing the engineered
non-Markovian reservoir in ultrasensitive detection.

\section{Acknowledgements}

We would like to thank Mr.~Wen-Lin Li for helpful discussions. This work was
supported by the NSF of China under Grant numbers 11474044 and 11547134.

\appendix

\section{Output signal of the mass sensor}

By using the general definition of the noise spectrum, according to Eq.\,(%
\ref{eqout}), we can obtain the output signal of the weak force detection
system
\begin{eqnarray}
S_{out}(\omega) &=& [\int d\omega ^{\prime }\langle
M_{out}(\omega)M_{out}(\omega ^{\prime })\rangle+\int d\omega ^{\prime
}\langle M_{out}(-\omega)M_{out}(-\omega ^{\prime })\rangle]/2  \nonumber \\
&=&\frac{|A(\omega)|^2+|B(\omega)|^2}{2}+S_{in}(\omega)|C(\omega)|^2
\end{eqnarray}
where $S_{in}(\omega)$ denotes the input signal from the external force $%
F_{ext}$ after ignoring the thermal noise of mechanical oscillator. Here we
intend to measure the mass of the human chromosome-1 as example, where the
external accretion mass will introduce an additional frequency responded by
mechanical resonator with mass responsivity $\mathcal{R}=10^{21}\textrm{Hz}%
\cdot \textrm{g}^{-1}$. Then, we deposit a few chromosomes onto the surface of
the mechanical resonator. The additional energy of the input signal can be
described as $S_{in}=Nm\mathcal{R}$, where $m=2.7 \times 10^{-13} \textrm{g}$
is the mass of one chromosome-1 molecule. $N$ is the number of the deposit
chromosomes. In the non-Markovian regime, $\omega_m$ is replaced by the
effective frequency $\omega_{eff}$, and the optimal detection area should be
on resonance with this effective frequency. We define the strength of the
frequency resonance spectrum $I_{out}=S_{out}(\omega_{eff})$, where $%
\omega_{eff}=\omega_{m}$ in the Markovian regime. The energy of $I_{out}$
will be linearly increases as the number of adsorbed chromosomes increases.
This allows us could detect the number of the chromosomes by measuring the
strength of the resonant output energy.

\section*{References}
\bibliography{omde}
\end{document}